\documentclass[conference]{IEEEtran}
\IEEEoverridecommandlockouts
\usepackage{cite}
\usepackage{amsmath,amssymb,amsfonts}
\usepackage{algorithmic}
\usepackage{graphicx}
\usepackage{textcomp}
\usepackage{url}
\usepackage{graphicx}
\usepackage{bm}
\usepackage{multirow}
\usepackage{booktabs}
\usepackage{makecell}
\usepackage{bbding}
\usepackage{balance}
\usepackage{float}
\usepackage{amsmath}
\usepackage{amsfonts}
\usepackage{bm}
\usepackage{color}
\usepackage{hyperref}
\usepackage{algorithmic}
\usepackage{algorithm}
\usepackage{mathrsfs}
\hypersetup{
    colorlinks=true,
    linkcolor=black,
    filecolor=black,      
    urlcolor=black,
    citecolor=black,
}
\usepackage{hyperref}
\usepackage[table]{xcolor}
\makeatletter
\def\UrlAlphabet{%
      \do\a\do\b\do\c\do\d\do\e\do\f\do\g\do\h\do\i\do\j%
      \do\k\do\l\do\m\do\n\do\o\do\p\do\q\do\r\do\s\do\t%
      \do\u\do\v\do\w\do\x\do\y\do\z\do\A\do\B\do\C\do\D%
      \do\E\do\F\do\G\do\H\do\I\do\J\do\K\do\L\do\M\do\N%
      \do\O\do\P\do\Q\do\R\do\S\do\T\do\U\do\V\do\W\do\X%
      \do\Y\do\Z}
\def\UrlDigits{\do\1\do\2\do\3\do\4\do\5\do\6\do\7\do\8\do\9\do\0}
\g@addto@macro{\UrlBreaks}{\UrlOrds}
\g@addto@macro{\UrlBreaks}{\UrlAlphabet}
\g@addto@macro{\UrlBreaks}{\UrlDigits}
\makeatother
\usepackage{threeparttable}
\usepackage{tabu}
\def\BibTeX{{\rm B\kern-.05em{\sc i\kern-.025em b}\kern-.08em
    T\kern-.1667em\lower.7ex\hbox{E}\kern-.125emX}}
\begin{document}

\title{Exploring Text-Queried Sound Event Detection with Audio Source Separation
% {\footnotesize \textsuperscript{*}Note: Sub-titles are not captured for https://ieeexplore.ieee.org  and
}

\author{\IEEEauthorblockN{
Han Yin$^{1,2}$, Jisheng Bai$^1$, Yang Xiao$^3$, Hui Wang$^2$, Siqi Zheng$^2$, Yafeng Chen$^2$,\\
Rohan Kumar Das$^3$, Chong Deng$^2$, Jianfeng Chen$^{1*}$\thanks{* Jianfeng Chen is corresponding author.}}
\IEEEauthorblockA{\textit{$^1$School of Marine Science and Technology, Northwestern Polytechnical University, Xi’an, China}}
\IEEEauthorblockA{\textit{$^{2}$Speech Lab, Alibaba Group, Hangzhou, China}~~~~\textit{$^{3}$Fortemedia Singapore, Singapore}}
% \IEEEauthorblockA{\textit{$^{4}$Centre for Vision, Speech and Signal Processing (CVSSP), University of Surrey, UK}}
% \IEEEauthorblockA{\textit{$^{4}$National University of Singapore, Singapore}}
}

% \author{\IEEEauthorblockN{1\textsuperscript{st} Han Yin}
% \IEEEauthorblockA{\textit{School of Marine Science and Technology} \\
% \textit{Northwestern Polytechnical University}\\
% Xi’an, China \\
% yinhan@mail.nwpu.edu.cn}
% \and
% \IEEEauthorblockN{2\textsuperscript{nd} Jianfeng Chen}
% \IEEEauthorblockA{\textit{School of Marine Science and Technology} \\
% \textit{Northwestern Polytechnical University}\\
% Xi’an, China \\
% chenjf@nwpu.edu.cn}
% \and
% \IEEEauthorblockN{3\textsuperscript{rd} Jisheng Bai}
% \IEEEauthorblockA{\textit{School of Marine Science and Technology} \\
% \textit{Northwestern Polytechnical University}\\
% Xi’an, China \\
% baijs@mail.nwpu.edu.cn}
% \and
% \IEEEauthorblockN{4\textsuperscript{th} Mou Wang}
% \IEEEauthorblockA{\textit{Institute of Acoustics} \\
% \textit{Chinese Academy of Sciences}\\
% Beijing, China \\
% wangmou21@mail.nwpu.edu.cn}
% \and
% \IEEEauthorblockN{5\textsuperscript{th} Mark D. Plumbley}
% \IEEEauthorblockA{\textit{ Centre for Vision, Speech and Signal Processing (CVSSP)} \\
% \textit{University of Surrey}\\
% Guilford GU2 7XH, U.K \\
% m.plumbley@surrey.ac.uk}
% % \and
% % \IEEEauthorblockN{6\textsuperscript{th} Given Name Surname}
% % \IEEEauthorblockA{\textit{dept. name of organization (of Aff.)} \\
% % \textit{name of organization (of Aff.)}\\
% % City, Country \\
% % email address or ORCID}
% }

\maketitle

\begin{abstract}
% 在声音事件检测(SED)任务中，重叠声音事件是一个很大的挑战，某些事件容易被淹没在背景噪声或者其他事件中，导致在这些事件上的检测性能很差。
% 为了解决这个问题，我们提出 queried SED (Q-SED) 框架。
% 具体地，我们首先使用一个预训练的language-based audio source separation (LASS) model，将声音事件文本作为queries，从输入音频中分离出不同事件对应的音轨。
% 然后，多个 target sound event detection (TSED) units 被用来对不同事件进行检测。
% 为了提升 LASS模型的性能，我们将 dual-path rnn block 融入 ResUNet 为了提取动态音频信息，并在大量的audio caption datasets上进行训练，最后所训练的模型在DCASE 2024 Task9 单模 track 上获得第一名。
% 实验结果表明，通过应用预训练的LASS模型，TQ-SED可以明显地提升SED性能，实现50.97\%的F1分数，超过传统的SED框架7.22\%。此外，TSED units的复杂度对检测性能的影响被充分探究。我们仅使用0.25 M的参数便可达到48.06\%的F1分数。代码被发布在xxxxxx.
In sound event detection (SED), overlapping sound events pose a significant challenge, as certain events can be easily masked by background noise or other events, resulting in poor detection performance. To address this issue, we propose the text-queried SED (TQ-SED) framework. Specifically, we first pre-train a language-queried audio source separation (LASS) model to separate the audio tracks corresponding to different events from the input audio. Then, multiple target SED branches are employed to detect individual events.
AudioSep is a state-of-the-art LASS model, but has limitations in extracting dynamic audio information because of its pure convolutional structure for separation.
To address this, we integrate a dual-path recurrent neural network block into the model. 
We refer to this structure as AudioSep-DP, which achieves the first place in DCASE 2024 Task 9 on language-queried audio source separation (objective single model track). 
Experimental results show that TQ-SED can significantly improve the SED performance, with an improvement of 7.22\%  on F1 score over the conventional framework. Additionally, we setup comprehensive experiments to explore the impact of model complexity. The source code and pre-trained model are released at \url{https://github.com/apple-yinhan/TQ-SED}.
\end{abstract}


\section{Introduction}
% Sound event detection (SED) involves identifying sound event classes and determining their timestamps, which has found widespread applications across various domains, such as smart cities \cite{urban-sed-1,urban-sed-2} and security systems \cite{security-sed-1}.

% In real-world scenarios, multiple events often occur simultaneously. Overlapping events significantly increase the difficulty of SED because the acoustic features of different events can mask or interfere with each other, making it challenging for the model to distinguish between them accurately, and leading to poor performance on certain event classes \cite{overlap, overlap2}.
% , and leading to poor detection performance for certain event classes. 
% especially those with weaker or less distinct acoustic signatures, as they can be overshadowed by more dominant sound events.
Sound event detection (SED) identifies sound event classes and their timestamps, and it has been widely applied in areas like smart cities~\cite{urban-sed-1,urban-sed-2} and security systems~\cite{security-sed-1}. A key benchmark in this field is the detection and classification of acoustic scenes and events (DCASE) challenge series, which includes a task focused on detecting sound events in domestic environments, highlighting the need for models that perform well in everyday settings. Advances in deep learning, particularly with convolutional recurrent neural networks (CRNN)~\cite{crnn4sed}, have greatly enhanced SED. A notable example is the frequency dynamic-CRNN~\cite{fdy}, which uses frequency-dependent kernels to improve event detection. However, these models primarily focus on designing novel structures for the acoustic model and overlook the challenge of classifying polyphonic sound events.
% , where multiple sounds occur.

In real-world environments, multiple events often occur at the same time. Overlapping events make SED more difficult because the acoustic features of different events can blend or interfere, making it hard for the model to separate them accurately~\cite{cyr2,dcase23-Xiao,dualkd,overlap, overlap2, cyr1, wildpaper}. This often leads to a poor performance in detecting specific event classes. Therefore, accurate detection of polyphonic sounds is critical for SED applications.
To tackle the challenges of overlapping events, one solution is to use audio source separation models. The authors of~\cite{sep_sed_1} used a sound separation network to isolate different sound events, but this network requires retraining when applied to new SED datasets. In the DCASE 2020 Challenge\footnote{\url{https://dcase.community/challenge2020/task-sound-event-detection-and-separation-in-domestic-environments}}, a universal sound separation model was used for SED, which was pre-trained on the free universal sound separation~\cite{fuss} dataset. This model can be directly applied to SED tasks without fine-tuning~\cite{sep_sed_2,sep_sed_3}, but it is limited in the types of sound sources it can separate. Recently, a new method called language-queried audio source separation (LASS) has emerged~\cite{lass_1, lass_2}. 
% The model pre-trained on large audio-text pairs, allows users to input text queries and separate the corresponding audio from a mixture, offering a ``separate anything you described'' capability. 
Unlike previous methods, LASS is not limited to specific sound source categories and can separate any event. Therefore, LASS has the potential to improve SED systems, but its effectiveness in this area has not been explored yet.

In this work, we propose the text-queried SED (TQ-SED) framework to explore the effectiveness of LASS models in SED tasks. Specifically, we propose a LASS model and pre-train it on large scale audio-text pairs. 
The proposed LASS model achieves the first place in the objective single model track of DCASE 2024 Task 9\footnote{DCASE 2024 Task 9: \url{https://dcase.community/challenge2024/task-language-queried-audio-source-separation-results}\label{dcase2024task9}} on language-queried audio source separation. 
In TQ-SED, the pre-trained LASS model is applied to separate target sound events, followed by multiple detection branches to identify each event. 
% We enhance the LASS model by incorporating a Dual-Path RNN (DPRNN)~\cite{dprnn} to improve separation performance and pre-train our model on large-scale audio caption datasets. 
The SED results indicate that applying the LASS model significantly improves detection performance for various sounds, especially for overlapping sound events. For instance, when event queries are applied for audio separation, the optimal threshod F1 scores for ``brakes squeaking'' and ``children voices'' increase by 26.71\% and 23.76\%, respectively.
% SED results show that applying LASS significantly improves the detection performance. 
The main contributions of this work are listed as follows:
\begin{itemize}
% \item We propose the TQ-SED framework, which uses a pre-trained LASS model for sound event separation, improving SED performance. This is the first study to explore text-queried SED with LASS models.
% \item To boost separation performance, we integrate a DPRNN block into ResUNet for dynamic audio extraction, and our model achieves state-of-the-art results on the DCASE 2024 evaluation dataset.
% % \item We analyze the impact of model complexity and find that even with only 0.25M learnable parameters, TQ-SED outperforms the baselines. We release the source code and pre-trained model to support future research.
\item We propose the novel TQ-SED framework using a pre-trained LASS model for sound event separation and enhance it with a dual-path recurrent neural network (DPRNN) block, achieving state-of-the-art (SOTA) results in the DCASE 2024 Task 9 evaluation dataset. {\it This is the first study on text-queried SED with LASS models.}
\item We show that TQ-SED outperforms baselines with only 0.25M learnable parameters. The source code and pre-trained model are released for future research.
\end{itemize}
\begin{figure*}
\centering
\centerline{\includegraphics[width=\textwidth]{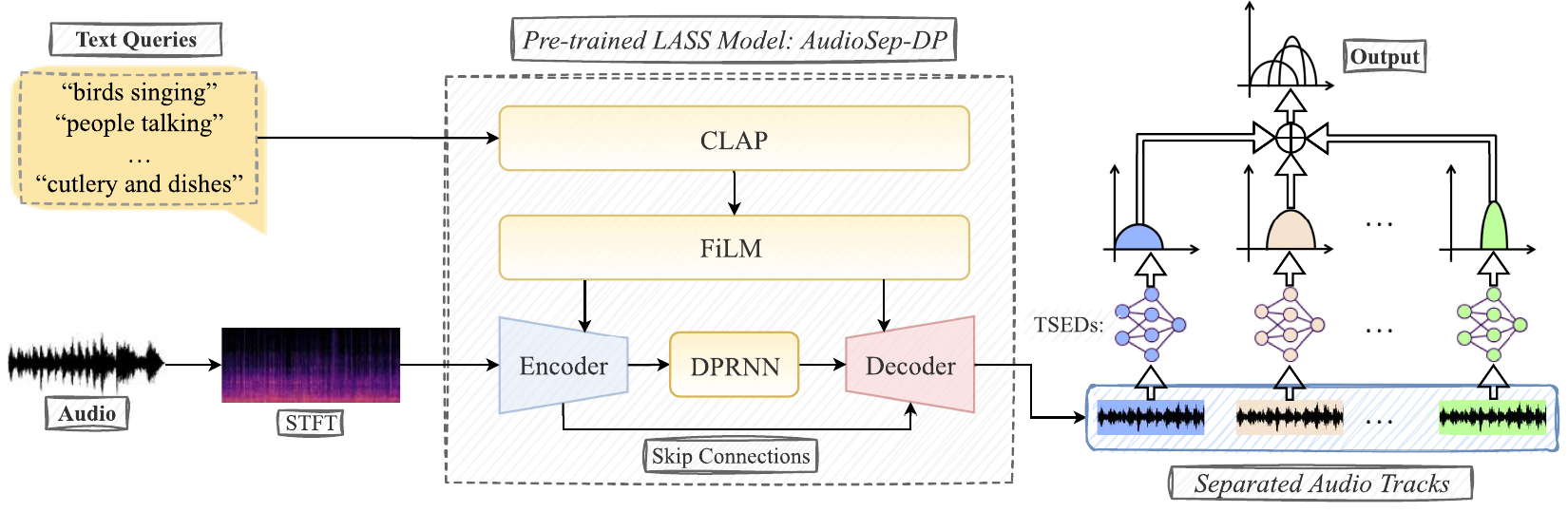}}
\caption{The overview of the proposed TQ-SED framework, where we pre-train AudioSep-DP on large scale audio-text pairs.}
\label{fig:overview}
\vspace{-2mm}
\end{figure*}

\section{Proposed Approach}
As shown in Fig.~\ref{fig:overview}, the proposed TQ-SED framework is mainly composed of two components: a pre-trained LASS model and multiple target sound event detection (TSED) branches. The LASS model is used for separating different sound event tracks based on text queries. It is noted that the LASS model is frozen during the whole process. We present a structure called AudioSep-DP for LASS, details about AudioSep-DP and TQ-SED are as follows.

\subsection{AudioSep-DP Pre-training}
AudioSep-DP comprises a text encoder and a separation network. Following \cite{lass_2}, we utilize the contrastive language-audio
pretraining (CLAP) \cite{clap} model as the text encoder and integrate the text embeddings into the separation network using feature-wise linear modulated (FiLM) layers \cite{film}.

For separation network, the frequency-domain ResUNet \cite{lass_1} is adopted in the SOTA system AudioSep \cite{lass_2}. However, as a purely convolutional structure, ResUNet has limitations in capturing dynamic audio information. To overcome this issue, we incorporate a DPRNN \cite{dprnn} block between the convolutional encoder and decoder of the ResUNet.
The DPRNN block is mainly composed of two bidirectional long short-term memory (Bi-LSTM) \cite{b-lstm} layers. 
The output of the convolutional encoder is denoted as $\bm{H}\in\mathbb{R}^{C\times T \times F}$, where $C$, $T$, and 
$F$ represent the number of channels, frames, and frequency bins, respectively. 
% First, 
% $\bm{H}$ is processed through a Bi-LSTM along the temporal dimension to capture dependencies between different frames. Subsequently, 
% $\bm{H}$ is fed into a second Bi-LSTM along the frequency dimension to extract the correlations between different frequency bins.
$\bm{H}$ is first processed through a Bi-LSTM along the temporal dimension to capture dependencies between frames,  then through another Bi-LSTM along the frequency dimension to extract correlations between frequency bins.
Following \cite{lass_1, lass_2}, we train an end-to-end LASS model using an L1 loss function between the predicted and the target waveforms, which can be formulated as:
\begin{equation}
    \text{Loss}_{L1} = \left|\hat{\bm{s}} - \bm{s}\right|
\end{equation}
where $\hat{\bm{s}}$ and $\bm{s}$ are the separated signal and the ground truth, respectively.

\subsection{Text-Queried Sound Event Detection}
In TQ-SED, we first use the pre-trained AudioSep-DP to separate audio for different events. Sound event labels are applied as the input queries. Let us consider the input audio as $\bm{x}$ and the text for the $i$-th event as $\bm{q}$, then the audio corresponding to the event can be obtained by: 
\begin{equation}
    \bm{z}_i = l_\theta(\bm{x}, \bm{q}), i=1,2,...,K
\end{equation}
where $l(\cdot)$ is the pre-trained AudioSep-DP, $K$ is the number of event classes, and $\theta$ is the pre-trained parameter.

Then, TSED branches are used to detect each event separately, as formulated in:
\begin{equation}
    \bm{y}_i = u_\alpha^i(\bm{z}_i), i=1,2,...,K
\end{equation}
where $u^i(\cdot)$ is the $i$-th TSED branch, and $\alpha$ is the learnable parameter. The final SED prediction $\bm{\hat{y}}\in\mathbb{R}^{T\times K}$ can be obtained by concatenating $\bm{y}_i\in\mathbb{R}^{T\times 1}$:
\begin{equation}
    \bm{\hat{y}} = \textrm{Concat}(\bm{y}_i), i=1,2,...,K
\end{equation}

In TQ-SED, each TSED branch only needs to perform binary classification, which is easier than multi-label classification. Therefore, we can use small-sized models, such as low-complexity CRNNs, to implement the TSED branches. Additionally, the TSED branches can be optimized simultaneously, with the loss computed on the final SED prediction. Since the SED dataset used in this work is soft-labeled with non-binary values, following \cite{dcase23-baseline}, we use the mean squared error (MSE) function to calculate the loss, which is more suitable than the binary cross-entropy loss, as formulated in:
\begin{equation}
    \text{Loss}_{\textrm{MSE}} = \Vert \bm{\hat{y}} - \bm{\tilde{y}} \Vert_2^2
\end{equation}
where $\bm{\tilde{y}}$ is the SED label. In this work, we use the CRNN model as the TSED branches, which mainly consists of three convolutional layers and one bidirectional gated recurrent unit (GRU) layer, and is also used as the baseline system in DCASE 2023 Task 4B\footnote{\url{https://dcase.community/challenge2023/task-sound-event-detection-with-soft-labels}\label{dcase23task4b}}. By adjusting the number of convolutional channels, the model size can be easily controlled.

\section{Experimental Setup}
In this section, we present details about the experimental setups for AudioSep-DP pre-training and text-queried SED.

\subsection{AudioSep-DP Pre-training}
\subsubsection{Datasets}
We use four public-released audio caption datasets as follows for training AudioSep-DP: DCASE-2024-T9Dev$^{\ref{dcase2024task9}}$,  Audiocaps \cite{audiocaps}, Auto-ACD \cite{auto-acd} and Wavcaps \cite{wavecaps}.

\textbf{DCASE-2024-T9Dev} is composed of audio samples from FSD50K \cite{fsd50k} and Clotho v2 \cite{clotho} datasets. Clotho v2 consists of 6972 audio samples, each audio clip is labeled with five captions. FSD50K contains over 51K audio clips manually labeled using 200 classes drawn from the AudioSet \cite{audioset} ontology.
\textbf{Audiocaps} is a large-scale dataset of 46K audio clips with human-written text pairs collected via crowdsourcing on the AudioSet dataset.
\textbf{Auto-ACD} contains over 1.9M audio-text pairs, which were generated based on a series of public tools or APIs, in which audio samples were from AudioSet and VGGSound \cite{vggsound}.
\textbf{Wavcaps} \cite{wavecaps} is the first large-scale weakly labeled audio captioning dataset, comprising approximately 400K audio clips with paired captions. Audio clips and their raw descriptions were from diverse sources, including FreeSound\footnote{\url{https://freesound.org}}, BBC Sound Effects\footnote{\url{https://sound-effects.bbcrewind.co.uk}}, SoundBible\footnote{\url{https://soundbible.com}}, and AudioSet. {\it Altogether, we use 1069 hours of data for training.}

For evaluation, we use data from DCASE 2024 challenge Task 9 evaluation (\textbf{DCASE-2024-T9EVal}) dataset$^{\ref{dcase2024task9}}$. 
The dataset contains 1000 audio clips, which are sampled from data uploaded to FreeSound between April and October 2023, and each clip has three captions.

% Finally, LASS systems are evaluated on blind evaluation dataset of DCASE 2024 challenge task 9 (DCASE-2024-T9Eval).
% This dataset consists of 1000 audio clips, which were generated in the same way as in DCASE-2024-T9Val.

\subsubsection{Evaluation Metrics}
Following \cite{lass_1, lass_2}, we use signal-to-distortion ratio (SDR), signal-to-distortion ratio improvement (SDRI) and scale-invariant signal-to-distortion ratio (SI-SDR) for evaluation.

\subsubsection{Training Strategy}
The sampling rate of audio samples in audio-text paired datasets is 16 kHz or 32 kHz. Therefore, we trained two LASS systems for each 16kHz and 32 kHz data.
Waveforms are converted to spectrograms using a short-time Fourier transform (STFT) with a window size of 0.064 s and a hop size of 0.01 s. We apply an Adam optimizer with a learning rate of 0.0001 to train the model. Batch size is set to 24, and all models are trained for 1 million steps.
We also train AudioSep \cite{lass_2} to explore the effectiveness of DPRNN.

\subsection{Text-Queried SED}
\subsubsection{Dataset}
We use a real-world polyphonic SED dataset, MAESTRO-Real\cite{maestro,dcase24-Xiao,dgpaper}, for training and validation. This dataset is also used in DCASE 2023$^{\ref{dcase23task4b}}$ / 2024\footnote{\url{https://dcase.community/challenge2024/task-sound-event-detection-with-heterogeneous-training-dataset-and-potentially-missing-labels}\label{dcase24task4}} challenges.
MAESTRO-Real consists of real-life audio clips recorded in 5 acoustic scenes: cafe restaurant, grocery store, city center, residential area, and metro station. There are 11 event classes to be evaluated: birds singing, car, people talking, footsteps, children voices, wind blowing,
brakes squeaking, large vehicle, cutlery and dishes, metro approaching and metro leaving. 
 
\subsubsection{Evaluation Metrics}
Micro-average error rate (ER) and macro-average F1 score are used to evaluate the SED performance.
When calculating ER and F1 scores, a decision threshold is usually used to determine whether an event occurs.
Ebbers et al. \cite{fscore} presented a methodology to calculate the optimal threshold for each event class.
In this work, F1 scores are all calculated with a class-specific optimal threshold.

\subsubsection{Training Strategy}
Following DCASE Challenge$^{\ref{dcase23task4b}}$, a 64-dimensional log-mel spectrogram is extracted with a window size of 400 ms and a hop size of 200 ms as the input acoustic feature
and models are obtained using a 5-fold cross-validation setup.
Adam optimizer is employed with a learning rate initialized to 0.001, which is automatically halved when there is no continuous performance improvement for 10 epochs.
Batch size is set to 32 and dropout rate is set to 0.2.
Training stops when the learning rate drops 10 consecutive times.

\subsubsection{Baselines}
To further demonstrate the effectiveness of the proposed TQ-SED framework, we present two SED frameworks as baselines to compare with ours.

\textbf{SED-Base-1} is the conventional SED framework, which has been used in most of the previous works \cite{crnn4sed}. In SED-Base-1, we directly input the audio mixture into a CRNN model for polyphonic event detection, meaning that the CRNN model is required to perform a multi-label classification task.

\textbf{SED-Base-2} is used to further validate the benefit of LASS. We remove the pre-trained LASS model from TQ-SED and directly input the audio mixture into TSED branches.

\section{Results and Discussions}
In this section, we discuss the results of LASS and SED studies. Details are described in the following subsections.
\subsection{Results of LASS}
% Table~\ref{tab:lass_dev} shows the results of different systems on DCASE-2024-T9Val.
% 32 kHz models perform better than 16 kHz models, with an improvement of 0.96 dB and 0.18 dB on SDR for ResUNet-based and DP-ResUNet-based models respectively.
% In addition, the proposed DP-ResUNet has a better separation performance than ResUNet, which demonstrates the effectiveness of the incorporated DPRNN block. 
% \begin{table}[htbp]
% \centering
% \caption{Results of LASS systems with different sampling rate and separation model on DCASE-2024-T9Val.}
% \renewcommand\arraystretch{1.5}{
% \setlength{\tabcolsep}{2.2mm}{
% \begin{tabular}{c|ccc}
% % \toprule[1pt]
% \hline
% Model   & SDR$\uparrow$ & SDRI$\uparrow$ & SISDR$\uparrow$\\
% % \hline
% %  ResUNet & 7.087 dB & 7.052 dB & 5.413 dB\\
% % DP-ResUNet & 8.007 dB & 7.972 dB & 6.459 dB \\
% \hline
% ResUNet & 8.047 dB & 8.012 dB & 6.558 dB\\
% DP-ResUNet & \textbf{8.191} dB & \textbf{8.156} dB & \textbf{6.794} dB \\

% \hline
% % \bottomrule[1pt]
% \end{tabular}
% % \begin{tablenotes}
% %    \footnotesize
% %    \item Note: $\star$ and $\ast$ represent 16 kHz and 32 kHz models, respectively.
% %    % \Hi{\item[$\dagger$] For closed source work, we built the network based on the description in the literature and estimated the parameters of the model.} 
% % \end{tablenotes}
% }
% }
% \label{tab:lass_dev}
% \end{table}

We present the separation performance of different systems on DCASE-2024-T9Eval in Table~\ref{tab:lass_eval}.
The proposed AudioSep-DP achieve the best separation performance at both 16 kHz and 32 kHz settings, outperforming the second-ranked system by 0.13 dB and 0.09 dB on SDR, respectively. In summary, our proposed 32 kHz AudioSep-DP model has the best performance among all SOTAs, with an SISDR of 7.349 dB.

\begin{table}[htbp]
\vspace{-1em}
\centering
\caption{Results of different LASS systems on DCASE-2024-T9Eval.}
\renewcommand\arraystretch{1.35}{
\setlength{\tabcolsep}{1.2mm}{
\begin{tabular}{c|c|ccc}
% \toprule[1pt]
\hline
 System & Sampling Rate  & SDR$\uparrow$ & SDRI$\uparrow$ & SISDR$\uparrow$\\
\hline
DCASE 2024 Baseline$^{\ref{dcase2024task9}}$ & 16 kHz & 5.799 dB & 5.693 dB & 3.873 dB\\
Xiao et al. \cite{dcase24-xiao}  & 16 kHz & 6.022 dB & 5.916 dB & 4.115 dB \\
Chuang et al. \cite{dcase24-chung} & 16 kHz& 7.302 dB & 7.195 dB & 5.628 dB \\
Romaniuk et al. \cite{dcase24-Romaniuk} & 16 kHz & 7.572 dB & 7.466 dB & 5.455 dB \\
Kim et al. \cite{dcase24-lee}&16 kHz & 8.059 dB & 7.953 dB & 6.510 dB \\
Xiao et al. \cite{dcase24-xiao}& 32 kHz & 8.368 dB & 8.262 dB & 6.800 dB \\
Kim et al.\cite{dcase24-lee} &32 kHz & 8.671 dB & 8.564 dB & 7.217 dB \\
\hline
AudioSep \cite{lass_2} & 16 kHz & 7.306 dB & 7.200 dB & 5.481 dB \\
AudioSep-DP & 16 kHz & 8.186 dB & 8.080 dB & 6.499 dB \\
AudioSep-DP & 32 kHz & \textbf{8.764} dB & \textbf{8.658} dB & \textbf{7.349} dB \\
% \hline
%% 32khz
% Guan et al. & 8.368 dB & 8.262 dB & 6.800 dB \\
% Kim et al. & 8.671 dB & 8.564 dB & 7.217 dB \\
% Ours & \textbf{8.764} dB & \textbf{8.658} dB & \textbf{7.349} dB \\
\hline
% \bottomrule[1pt]
\end{tabular}
% \begin{tablenotes}
%    \footnotesize
%    \item Note: $\star$ and $\ast$ represent 16 kHz and 32 kHz models, respectively.
%    % \Hi{\item[$\dagger$] For closed source work, we built the network based on the description in the literature and estimated the parameters of the model.} 
% \end{tablenotes}
}
}
\label{tab:lass_eval}
\end{table}

\subsection{Results of Text-Queried SED}
In TQ-SED, we use the pre-trained AudioSep-DP model to separate audio based on text queries, which is frozen during the whole process. Event labels are applied as the input queries. Table~\ref{tab:sed_1} shows the SED performance and total learnable parameters of different frameworks on the MAESTRO-Real dataset. By gradually reducing the number of convolutional filters in CRNN, we can gradually reduce model parameters. The proposed TQ-SED framework significantly outperforms both SED-Base-1 and SED-Base-2, which illustrates the effectiveness of the pre-trained AudioSep-DP. In addition, when the parameters drop from 4.18 M to 0.25 M, the F1 score drops from 50.97\% to 48.69\%. However, this is still 4.17\% higher than SED-Base-2 with 4.18 M parameters.
\begin{table}[t]
\centering
\caption{Results of different frameworks on MAESTRO-Real.}
\renewcommand\arraystretch{1.3}{
\setlength{\tabcolsep}{3mm}{
\begin{tabular}{c|c|c|cc}
% \toprule[1pt]
\hline
Framework& Conv Filters   & Params & ER$\downarrow$ & F1$\uparrow$\\
\hline
SED-Base-1 & 128 & 0.38 M  & 0.472 & 43.75\%\\
\hline
SED-Base-2 & 128 & 4.18 M  & 0.463 & 44.52\%\\
\hline
\multirow{4}{*}{\makecell{TQ-SED\\(Proposed)}} & 128 & 4.18 M & \textbf{0.425} & \textbf{50.97}\%\\
& 64 & 1.32 M& 0.446 & 49.39\% \\
& 32 & 0.51 M& 0.455 & 48.73\% \\
& 16 & 0.25 M& 0.453 & 48.69\% \\
\hline
% \bottomrule[1pt]
\end{tabular}
% \begin{tablenotes}
%    \footnotesize
%    \item Note: $\star$ and $\ast$ represent 16 kHz and 32 kHz models, respectively.
%    % \Hi{\item[$\dagger$] For closed source work, we built the network based on the description in the literature and estimated the parameters of the model.} 
% \end{tablenotes}
}
}
\label{tab:sed_1}
\end{table}

\begin{table}[t]
\vspace{-1em}
\centering
\caption{Results of TQ-SED with different pre-trained LASS models on MAESTRO-Real.}
\renewcommand\arraystretch{1.3}{
\setlength{\tabcolsep}{2.2mm}{
\begin{tabular}{c|c|cc}
% \toprule[1pt]
\hline
\multirow{2}{*}{LASS Model} & Audio Separation& \multicolumn{2}{c}{SED}\\
& SDR$\uparrow$& ER$\downarrow$ & F1$\uparrow$\\
\hline
DCASE 2024 Baseline$^{\ref{dcase2024task9}}$ & 5.799 dB & 0.461 & 45.81\%\\
AudioSep-DP (16 kHz) & 8.186 dB & 0.459 & 46.23\%\\
AudioSep-DP (32 kHz) & \textbf{8.764} dB&\textbf{0.425} & \textbf{50.97}\%\\ 
\hline
% \bottomrule[1pt]
\end{tabular}
}
}
\label{tab:sed_2}
\end{table}

\begin{figure}[t]
\centering
\centerline{\includegraphics[width=0.5\textwidth]{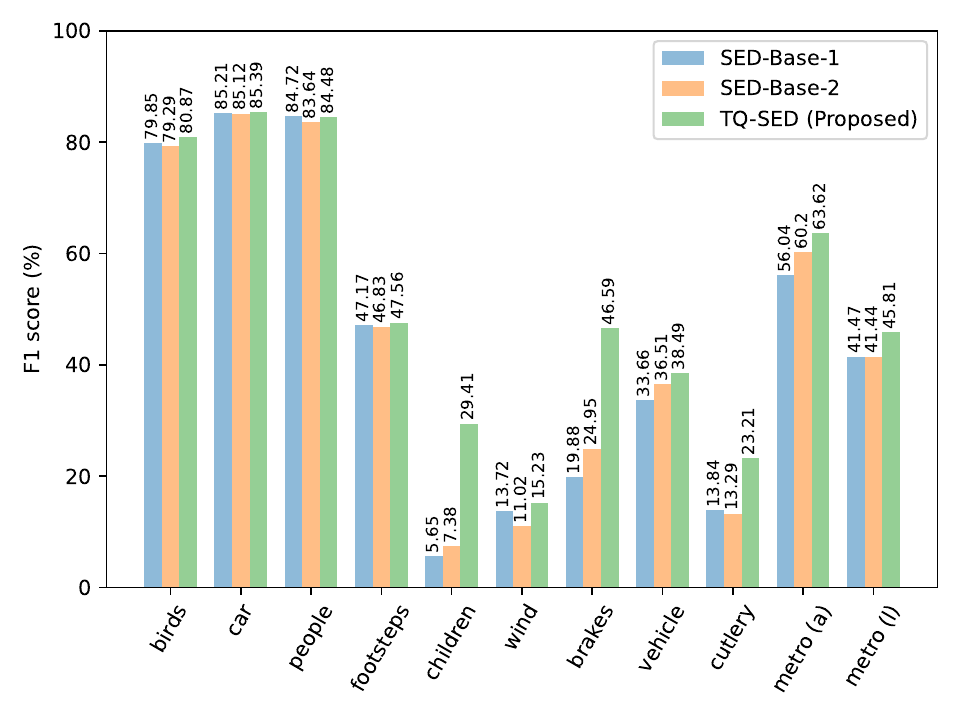}}
\vspace{-2mm}
\caption{F1 scores for each event of different frameworks on MAESTRO-Real.}
\label{fig:score_each_event}
\vspace{-1em}
\end{figure}

\begin{table}[t]
\centering
\caption{The Ratio of overlapping events in MAESTRO-Real.}
\renewcommand\arraystretch{1.3}{
\setlength{\tabcolsep}{1.3mm}{
\begin{tabular}{c|cccc|c}
% \toprule[1pt]
\hline
\multirow{2}{*}{Event} & \multicolumn{4}{c|}{Number of Co-occurring Events}   & \multirow{2}{*}{Total Duration}\\
 & 0 & 1 & 2 & 3 & \\
 \hline
birds singing & 39.74\% &	54.28\% &	5.23\% &	0.75\%  & 36.80 min\\
car	 & 38.29\% & 	46.63\% & 	13.21\% & 	1.88\% & 	87.83 min\\
people talking	& 76.68\% & 	14.68\% & 	7.57\% & 	1.07\% & 	142.58 min\\
footsteps	& 23.89\% & 	43.07\% & 	28.39\% & 	4.64\% & 	 36.10 min \\
children voices	& 19.13\% &	43.72\% & 	25.68\% & 	11.48\% & 	 4.58 min \\
wind blowing & 9.93\% & 85.82\% & 	3.55\% & 	0.71\% & 	 3.53 min \\
brakes squeaking & 0.00\% &  61.82\% & 	23.64\% & 	14.55\% & 	2.75 min \\
large vehicle& 	15.38\% & 	69.23\% & 	12.05\% & 	3.34\% & 	17.23 min \\
cutlery and dishes&	12.94\% & 	82.94\% & 	4.12\% & 	0.00\% & 	4.25 min\\
metro approaching&	75.22\% & 	23.23\% & 	1.33\% & 	0.22\% & 	11.30 min \\
metro leaving&	70.57\% & 	27.42\% & 	1.67\% & 	0.33\% & 	7.48 min \\
\hline
% \bottomrule[1pt]
\end{tabular}
}
}
\label{tab:sed_overlap}
\vspace{-2mm}
\end{table}

In Table~\ref{tab:sed_2}, we present the SED performance of TQ-SED with different pre-trained LASS models on the MAESTRO-Real dataset. It is observed that as the separation performance of the LASS model improves, the detection performance gradually increases, which verifies the importance of audio source separation to the SED performance.

Fig.~\ref{fig:score_each_event} presents the F1 score of each event for different frameworks. 
TQ-SED has superior detection performance on all events compared to both baselines. To evaluate the effectiveness of TQ-SED in handling overlapping events, we present the overlap statistics for each event in Table~\ref{tab:sed_overlap}. As shown, the maximum number of overlapping events is 4. Certain events, such as ``children voices'' and ``brakes squeaking'', rarely occur independently and are usually accompanied by other events. We find that TQ-SED demonstrates significant improvement in detecting these frequently overlapping events, notably achieving a 26.71\% increase in the F1 score for ``brakes squeaking'' compared to the conventional framework SED-Base-1. Finally, we present the performance of some other SOTA systems in Table~\ref{tab:sed_3}, which also shows that TQ-SED has a superior performance in terms of ER and F1.

% especially on the events ``children voices'', ``brakes squeaking'' and ``cutlery and dishes''. 
% The F-scores of TQ-SED on these three events are 23.76\%, 26.71\% and 9.37\% higher than those of the traditional SED-Base-1 framework, respectively.
% We observe that the sounds associated with these three types of events are typically transient and sporadic, often occurring in conjunction with continuous events, such as ``people talking'' and ``car'', rather than in isolation.
% Fig.~x shows a sample where the pre-trained LASS model is used to separate the event ``cutlery and dishes'' from the mixture.
% When LASS is not used for audio separation, i.e., in SED-Base-1 and SED-Base-2, the F1 scores for the events ``children voices'', ``wind blowing'', ``brakes squeaking'', ``large vehicle'', and ``dishes and cutlery'' are significantly lower than those of other events. However, in TQ-SED, when LASS is incorporated, the F1 scores for these events improve substantially, indicating that LASS can effectively separate these events from the original audio, enhancing the SED performance.
% The results show that the proposed system outperforms most SOTAs in terms of both ER and F1. 

\begin{table}[t!]
\centering
\caption{Results of different SOTA systems on MAESTRO-Real.}
\renewcommand\arraystretch{1.3}{
\setlength{\tabcolsep}{8mm}{
\begin{tabular}{c|ccc}
% \toprule[1pt]
\hline
System    & ER$\downarrow$ & F1$\uparrow$\\
\hline
Zhang et al. \cite{dcase23-Zhang}    & 0.443 & 44.49\%\\
Tri-Do et al. \cite{dcase23-Tri-Do}    & 0.450 & 46.71\%\\
Min et al. \cite{dcase23-Min}   & 0.445 & 45.81\%\\
Chen et al. \cite{dcase23-Chen}    & 0.430 & 49.70\%\\
\hline
\makecell{Proposed}  & \textbf{0.425} & \textbf{50.97}\%\\
% & 0.446 & 49.39\% \\
% & 0.455 & 48.43\% \\
% & 0.461 & 48.06\% \\
\hline
% \bottomrule[1pt]
\end{tabular}
% \begin{tablenotes}
%    \footnotesize
%    \item Note: $\star$ and $\ast$ represent 16 kHz and 32 kHz models, respectively.
%    % \Hi{\item[$\dagger$] For closed source work, we built the network based on the description in the literature and estimated the parameters of the model.} 
% \end{tablenotes}
}
}
\label{tab:sed_3}
\vspace{-1em}
\end{table}

\section{Conclusions}
In this work, we introduce the novel TQ-SED framework, which leverages the advantages of the LASS model for SED. The proposed AudioSep-DP, pre-trained on large-scale audio-text pairs, is integrated into the TQ-SED framework to extract audio tracks corresponding to distinct sound events. 
% Multiple TSED branches are then applied to detect each event separately. 
Our proposed AudioSep-DP model surpasses all SOTAs on the DCASE-2024-T9EVal benchmark according to objective metrics. Experimental results demonstrate that using the pre-trained AudioSep-DP to separate audio tracks for different sound events significantly enhances overall SED performance, especially for challenging event categories. We hope this work offers valuable insights for the SED domain.
% In this paper, we propose the TQ-SED framework, to explore the potential benefits of using a LASS model for SED, where the LASS model is pre-trained on large-scale audio-text pairs. 
% Specifically, the TQ-SED framework first uses the pre-trained LASS model to extract audio tracks of different sound events, and then applies multiple TSED branches to detect each event separately.
% The proposed LASS model outperforms all SOTAs on DCASE-2024-T9EVal in terms of objective metrics.
% Experimental results demonstrate that separating audio of different sound events using the pre-trained LASS model can enhance the overall performance of SED, especially for some hard-to-detect event categories. We hope this work provides insights for the SED domain, and a promising future research direction would be to develop a unified model that simultaneously performs both audio source separation and event detection.

\balance

% \vspace{12pt}
% \color{red}
% IEEE conference templates contain guidance text for composing and formatting conference papers. Please ensure that all template text is removed from your conference paper prior to submission to the conference. Failure to remove the template text from your paper may result in your paper not being published.

\end{document}